\documentclass{article}
            \usepackage{graphics}
            \usepackage{amsmath,amssymb}
            \begin{document}
            \title{New Gauge Fields from Extension of
            Parallel Transport of
            Vector Spaces to  Underlying Scalar Fields}
            \author{Paul Benioff,\\
            Physics Division, Argonne National
            Laboratory,\\ Argonne, IL 60439, USA \\
            e-mail:pbenioff@anl.gov}

            \maketitle

            \begin{abstract}
            Gauge theories can be described by assigning a vector space $\bar{V}(x)$ to each space time point $x.$ A common set of complex numbers, $\bar{C},$  is usually assumed to be the set of scalars for all the $\bar{V}_{x}$.  This is expanded here to assign a separate set of scalars, $\bar{C}_{x}$, to $\bar{V}_{x}.$ The freedom of choice of bases, expressed by the action of a gauge group operator on the $\bar{V}_{x}$, is expanded here to include the freedom of choice of scale factors, $c_{y,x},$ as elements of $GL(1,C)$ that relate $\bar{C}_{y}$ to $\bar{C}_{x}.$ A gauge field representation of $c_{y,x}$ gives two gauge fields, $\vec{A}(x)$ and $i\vec{B}(x).$ Inclusion of these fields in the covariant derivatives of Lagrangians results in $\vec{A}(x)$ appearing as a gauge boson for which mass is optional and $\vec{B}(x)$ as a massless gauge boson. $\vec{B}(x)$ appears to be the photon field.  The nature of $\vec{A}(x)$ is not known at present. One does know that the coupling constant of $\vec{A}(x)$ to matter fields is very small compared to the fine structure constant.
            \end{abstract}


            \section{Introduction}
           The assignment of different vector spaces to different
            space time points has been used as a framework to describe
            some physical theories.  This
            approach and the freedom to choose bases in the
            different spaces \cite{Yang} has resulted in the
            development of several different gauge theories,
            such as QED and QCD. They also play an important
            role in the standard model \cite{Novaes}.

            This approach to gauge theories is based on the use of
            one common complex number field, $\bar{C}$, as the set
            of scalars for the vector spaces at different points.  All vector
            space operations involving scalars have scalar values in
            $\bar{C}.$

            Here a different approach is used in which a separate complex
            number field, $\bar{C}_{x},$ is associated with a
            vector space, $\bar{V}_{x},$ at each space time point $x.$
            The scalars in scalar-vector multiplication and scalar
            products of vectors in $\bar{H}_{x}$ take values in
            $\bar{C}_{x}.$ In the following, vector spaces
            will be limited to be Hilbert spaces.

            Some consequences of this expansion of the usual setup
            are explored here. The presence of different scalar fields at each point makes it possible to extend the freedom of choice of
            basis sets in each $\bar{H}_{x}$ \cite{Yang}
            to include freedom of choice of  complex number structures $\bar{C}_{x}$ that differ from one another by scaling
            factors \cite{BenNGF,BenRENT}.

            A good place to begin is with a description of parallel transformations between Hilbert spaces.  Here
            the use of unitary parallel transform  operators
            $U_{y,x}$ to map $\bar{H}_{x}$ onto $\bar{H}_{y}$
            \cite{Mack,Montvay} is expanded to include parallel
            transform operators $F_{y,x}$ to map $\bar{C}_{x}$ onto
            $\bar{C}_{y}.$  Both these operators define what is
            meant by the same vector and same number value.
            If $\psi_{y}$ and $\psi_{x}$ are vectors in $\bar{H}_{y}$
            and $\bar{H}_{x}$, then $\psi_{y}=U_{y,x}\psi_{x}$ is the same vector in $\bar{H}_{y}$ as $\psi_{x}$ is in $\bar{H}_{x}.$
            $a_{y}=F_{y,x}a_{x}$ is the same number value in $\bar{C}_{y}$ as $a_{x}$ is in $\bar{C}_{x}.$

            If $U_{y,x}$ and $F_{y,x}$ are to include the freedom of choice of bases and of scaling factors, then these operators must each be factored into the product of two operators. This follows from the fact that $U_{y,x}$ cannot be represented by a matrix of numbers or used of elements of a Lie algebra. Similarly $F_{y,x}$ cannot be represented by a analytic function.  This is shown in detail in the next section.

            The rest of the paper is devoted to exploring
            consequences of scaling of the complex number fields.
            The description will be brief as details have been given for
            complex (and other types of) numbers elsewhere
            \cite{BenRENT}. Also this paper expands on an earlier
            treatment where the scaling factors were restricted to
            be real numbers \cite{BenNGF}.

            \section{Factorization of Parallel
            Transforms}\label{FPT}Factoring unitary
            parallel transform operators in quantum theory is
            necessary if one uses the usual representations of
            unitary operators.  However it is not done in practice
            as it leads to nothing new in the usual treatments. However,
            for the purposes of this work it is useful to understand the problem as factorization is needed.

            Let $\bar{H}_{x}$ and $\bar{H}_{y}$ be two $n$ dimensional
            Hilbert spaces at space time points $x,y$, and $U_{y,x}$ a
            unitary operator  from $\bar{H}_{x}$ onto $\bar{H}_{y}.$  As an element of the gauge group, $U_{y,x}$ is supposed
            to account for the freedom of basis choice \cite{Yang}
            between $\bar{H}_{x}$ and $\bar{H}_{y}.$

            A problem arises if one attempts to represent $U_{y,x}$
            as a matrix of numbers or as the exponential of Lie algebra operators. If $U_{y,x}$ is so represented, then the action of $U_{y,x}$ on a vector in $\bar{H}_{x}$ is a vector in $\bar{H}_{x}.$  It is not a vector in $\bar{H}_{y}.$

            This problem can be solved by representing $U_{y,x}$
            as the product of two unitary operators: \begin{equation}\label{UXV}U_{y,x}=X_{y,x}V_{y,x}.
            \end{equation}$V_{y,x}:\bar{H}_{x}
            \rightarrow\bar{H}_{x}$ is a map from $\bar{H}_{x}$ to
            $\bar{H}_{y}$,  and $X_{y,x}:\bar{H}_{x}\rightarrow\bar{H}_{y}$
            is a map from $\bar{H}_{x}$ to $\bar{H}_{y}.$ If $\psi$ is a field with vector value, $\psi(y),$ in $\bar{H}_{y}$, then $V_{y,x}\psi(y)_{x}=
            X^{\dag}_{y,x}\psi(y)$ is the local representation of
            $\psi(y)$ in $\bar{H}_{x}.$ Here $U_{y,x}$ is a parallel transformation operator from $\bar{H}_{x}$ to $\bar{H}_{y}$  that defines the notion of sameness between the two vector spaces. $\psi(y)_{x}= U^{\dag}_{y,x} \psi(y)$ is the same vector in $\bar{H}_{x}$ as $\psi(y)$ is in $\bar{H}_{y}.$ Here $V_{y,x}$ can be represented by a matrix of numbers or through use of Lie algebra operators.  The fact that  $X_{y,x}$ cannot be so represented is now not a problem.

            If one expands the freedom of basis choice to include the freedom of choice of complex numbers as scalars, then similar problems exist for the complex numbers.
            As was the case for vector spaces, these problems can be solved by
            describing a local representation of $\bar{C}_{y}$ on $\bar{C}_{x}.$ This can be done by factoring  the parallel transformation operator
            $F_{y,x}:\bar{C}_{x}\rightarrow\bar{C}_{y}$  into two operators:
            \begin{equation}\label{FWW}F_{y,x}=W^{y}_{c}W^{c}_{x}.
            \end{equation} Both $W^{y}_{c}$ and $W^{c}_{x}$ are
            isomorphisms  with $W^{c}_{x}$ a map from $\bar{C}_{x}$
            onto $\bar{C}_{x}$ and $W^{y}_{c}$ a map from $\bar{C}_{x}$
            onto $\bar{C}_{y}.$ If $a_{y}$ is a number in $\bar{C}_{x}$ and
            $a_{x}=F^{-1}_{y,x}a_{y}$ is the same number in $\bar{C}_{x}$
            as $a_{y}$ is in $\bar{C}_{y},$ then $W^{c}_{x}a_{x}=
            (W^{y}_{c})^{-1}a_{y}$ is the representation of $a_{y}$ in
            $\bar{C}_{x}.$ One can extend this to the complex number fields
            and define the local representation of $\bar{C}_{y}$ on $\bar{C}_{x}$
            by $W^{c}_{x}\bar{C}_{x}=(W^{y}_{c})^{-1}\bar{C}_{y}.$

            \section{Digression}\label {D}
            Here some material is presented to help make the
            material in the next section easier to understand. The  mathematical logical description \cite{Barwise,Keisler} of different types of mathematical systems as structures is used here.  A structure consists of a base set, $1$ or more basic operations, $0$ or
            more basic relations, and $1$ or more constants.  The
            structures are required to satisfy a set of
            axioms appropriate for the system type being considered \cite{Keisler}.

            For example, a complex number structure  is given by
            \begin{equation}\label{Cbar}\bar{C}=\{C,+,-,\times,\div,0,1\}.
            \end{equation}$C$ with an overline denotes a structure. Without
            an overline, it denotes a base set. The axioms that $\bar{C}$
            must satisfy are those of an algebraically closed field of
            characteristic $0$ \cite{complex}.

            The material in the next section, which is the main part of this
            paper, is based on the discovery that it is  possible to define
            complex number structures (and structures of any number type)
            that differ from one another by arbitrary complex scaling factors.
            For each complex number, $c,$ one can define a structure,
            $\bar{C}^{c},$ on $C$ in which the number value, $c,$ in $\bar{C}$
            is the identity in $\bar{C}^{c}.$  This scaling of number values
            between $\bar{C}^{c}$ and $\bar{C}$ requires compensatory scaling
            of  the basic operations in $\bar{C}^{c}$ in
            terms of those in $\bar{C}.$  The compensatory scaling must be
            such that $\bar{C}^{c}$ satisfies the complex number axioms
            if and only if $\bar{C}$ does.

            A very simple example of this scaling is quite useful to help in
            understanding the scaling.  Let $\bar{N}$ be a structure,
            \begin{equation}\label{Nbar}\bar{N}=\{N,+,\times,<,0,1\},
            \end{equation}for the natural numbers $0,1,2,3,\cdots.$ $\bar{N}$
            satisfies the axioms \cite{Kaye} for the natural numbers.

            Consider the even numbers $0,2,4,\cdots.$ One would expect these
            to also be a valid model for the natural number axioms.
            Here $2$ plays the role of the identity. The corresponding structure
            can be represented by \begin{equation}\label{barN2}
            \bar{N}^{2}=\{N_{2},+_{2},\times_{2},0,1_{2}\}.\end{equation}
            This is a structure in which any even number, $2n,$ in $\bar{N}$ is
            assigned the value $n$ in $\bar{N}^{2}.$

            However, the goal is to give another representation of $\bar{N}^{2}$
            in terms of the basic operations and number valuations in $\bar{N}$.
             This means that the number value $2,$ in $\bar{N}$ must have the
             properties of the identity in the other representation of $\bar{N}^{2}.$ This seems impossible as $2$ is not $1.$ Yet it is possible if one realizes that the axiomatic definition of the number $1$ is that it is the multiplicative identity.

             It follows that the number value $2$ can serve as the multiplicative
             identity, if one changes the definition of multiplication in
             $\bar{N}^{2}$ to reflect the scaling. The structure $\bar{N}^{2}$
             can now be written as \begin{equation}\label{N2bar}
             \bar{N}^{2}=\{N_{2},+,\frac{\times}{2},<,0,2.\}.\end{equation}
             This shows that addition in $\bar{N}^{2}$ is the same as that
             in $\bar{N}$, but multiplication has changed in that a factor
             of $2$ has been included.

             The proof that, with this definition of multiplication, the
             number value, $2$ is the multiplicative identity in $\bar{N}^{2}$,
             follows from the equivalences: $$n_{2}\times_{2}1_{2}=n_{2}
             \Leftrightarrow 2n\frac{\textstyle \times}{\textstyle 2}2=
             2n\Leftrightarrow n\times 1=n.$$ The first equation is in
             $\bar{N}^{2},$ Eq. \ref{barN2}, the second is in $\bar{N}^{2},$
             Eq. \ref{N2bar}, and the third is in $\bar{N}.$  These equivalences
             show that, as required, $2$ is the multiplicative identity in
             $\bar{N}^{2}$ if and only if $1$ is the multiplicative identity
             in $\bar{N}.$

             These ideas are applied in the next section to complex number
             structures where scaling is by a complex number $c$ that depends on
             space and time.  Representation of $\bar{C}_{y}$ on $\bar{C}_{x}$
             correspond to the descriptions of both representations of
             $\bar{N}^{2}$ relative to that of $\bar{N}.$

            \section{The Representation of $\bar{C}_{y}$ on $\bar{C}_{x}$}
            \label{RCyCx}In this work, complex number structures are associated with each space time point. For points $x,y$ the structures,
            $\bar{C}_{y}$ and $\bar{C}_{x}$ are given by
            \begin{equation}\label{bCybCx}\begin{array}{c}\bar{C}_{y}=\{C_{y},
            +_{y},-_{y},\times_{y},\div_{y},^{*_{y}},0_{y},1_{y}\}
            \\\bar{C}_{x} =\{C_{x},+_{x},-_{x},\times_{x},\div_{x},^{*_{x}},
            0_{x},1_{x}\}.\end{array}\end{equation} Here $C_{y}$ and $C_{x}$, without over lines are the base sets of the structures, $+,-,\times,\div$ are the basic operations,
            and $0,1$ are constants. The complex conjugation operation,
            $\mbox{}^{*}$ has been added as it simplifies the development. The
            subscripts denote structure membership of the operations and constants. Both $\bar{C}_{x}$ and $\bar{C}_{y}$ satisfy the axioms for complex numbers \cite{Barwise,Keisler}.

            The structure $\bar{C}^{c}_{x}$ where \begin{equation}\label{bCcx}
            \bar{C}^{c}_{x}=\{C_{x},+_{c},-_{c},\times_{c},
            \div_{c},^{*_{c}},0_{c},1_{c}\}
            \end{equation}is defined to be the local representation of
            $\bar{C}_{y}$ at $x.$ Here the base set, is denoted by $C_{x}$ as
            it is the same set as that in $\bar{C}_{x}.$ The relations between
            $\bar{C}_{x},\bar{C}^{c}_{x},$ and $\bar{C}_{y}$ are given by
            \begin{equation}\label{CyWycWcxCx}\bar{C}_{y}=W^{y}_{c}\bar{C}^{c}_{x}=
            W^{y}_{c}W^{c}_{x}\bar{C}_{x}. \end{equation}

            It remains to give the explicit description of the structure
            elements of $\bar{C}^{c}_{x}$ in terms of those in
            $\bar{C}_{x}.$ Let $y=x+\hat{\nu}dx$ be a neighbor point
            of $x.$ The isomorphism $W^{c}_{x}$ is given by
            \begin{equation}\label{Wcx}
            \begin{array}{c}W^{c}_{x}(a_{x})= ca_{x},\\\\ W^{c}_{x}(\pm_{x})
            = \pm_{x},\hspace{1cm}W^{c}_{x}(\times_{x})= \frac{\textstyle\times_{x}}
            {\textstyle c},\\\\W^{c}_{x}(\div_{x}) = c\div_{x},\hspace{1cm}
            W^{c}_{x}((a_{x})^{*_{x}})= c(a^{*_{x}}_{x}).\end{array}\end{equation}

            From this one can describe $\bar{C}^{c}_{x}$ explicitly by
            \begin{equation}\label{Ccx}\bar{C}^{c}_{x}=\{C_{x},+_{x},-_{x},
            \frac{\times_{x}}{c},c\div_{x},c(-)^{*_{x}},0_{x},1_{x}\}.
            \end{equation} Here $c=c_{y,x}$ is a complex number in
            $\bar{C}_{x}$ that is associated with the link from $x$
            to $y.$ Also, the base set $C_{x}$ is
            the same in both $\bar{C}^{c}_{x}$ and in $\bar{C}_{x}.$

            Comparison of number values in $\bar{C}^{c}_{x}$ and $\bar{C}_{x}$
            shows that that they are scaled by the factor $c.$ A number
            value $a_{c}$ in $\bar{C}^{c}_{x}$ corresponds to the number
            value $ca_{x}$ in $\bar{C}_{x}.$ Here $a_{c}$ is the same number
            value in $\bar{C}^{c}_{x}$ as $a_{x}$ is in $\bar{C}_{x}.$

            One sees from these relations that "correspondence" is distinct
            from "sameness." The number value in $\bar{C}_{x}$ that corresponds
            to $a_{c}$ in $\bar{C}^{c}_{x}$ is different from the number
            value in $\bar{C}_{x}$ that is the same  as $a_{c}$ is in $\bar{C}^{c}_{x}.$ These two concepts coincide if and only
            if $c=1$. This describes the usual case where the $\bar{C}_{x}$ are all the same and can be replaced by one $\bar{C}.$

            Note that $a_{c}^{*_{c}}=(ca_{x})^{*_{c}}$ corresponds to
            $c(a_{x})^{*_{x}}.$ It does \emph{not} correspond to
            $c^{*_{x}}a_{x}^{*_{x}}.$ This follows from the
            equivalences $$\begin{array}{l}1_{c}^{*_{c}}=1_{c}\Leftrightarrow
            (c1_{x})^{*_{c}}=c1_{x}\\\hspace{1cm}\Leftrightarrow
            c(1_{x}^{*_{x}})=c1_{x}\Leftrightarrow
            1_{x}^{*_{x}}=1_{x}.\end{array}$$

            Another aspect of the relation between $\bar{C}^{c}_{x}$ and $\bar{C}_{x}$ is that one must drop the usual assumption that the elements of the base set, $C_{x},$ have fixed values, independent of the structure containing the base set.  Here the number values associated with the elements of $C_{x},$ with one exception, depend on the structure containing $C_{x}.$ The element
            of $C_{x}$ that has value $a_{c}$ in $\bar{C}^{c}_{x}$ has value $ca_{x}$ in $\bar{C}_{x}.$ This is different from the element of $\bar{C}_{x}$ that has the same value, $a_{x},$ as $a_{c}$ is in $\bar{C}^{c}_{x}.$

\begin{figure}[h]\begin{center}
           \resizebox{100pt}{100pt}{\includegraphics[250pt,200pt]
           [500pt,500pt]{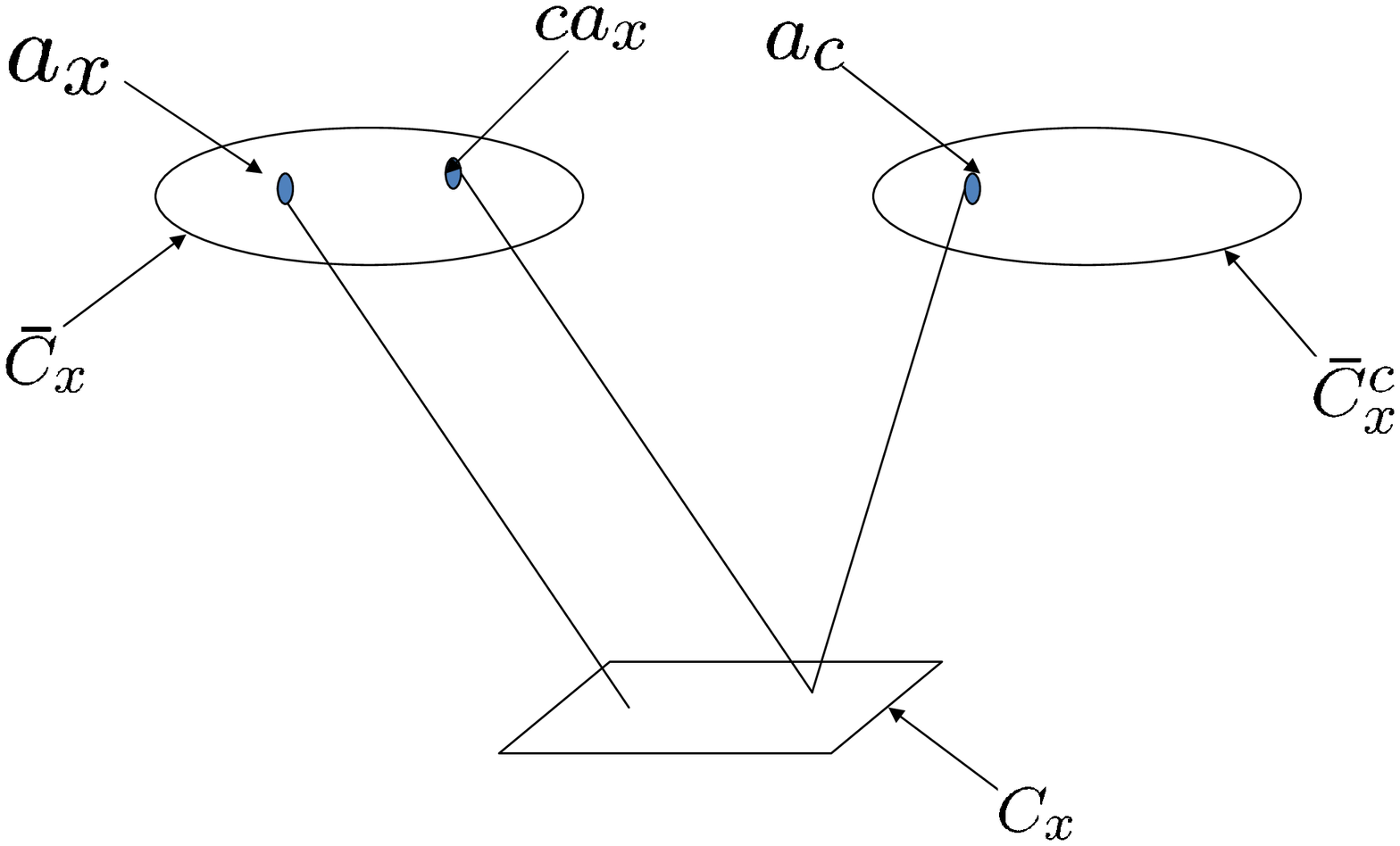}}\end{center}
           \caption{Relations between Elements in the base set $C_{x}$
           and their Numerical Values in  the Structures $\bar{C}_{x}$
           and $\bar{C}^{c}_{x}.$ Here $a_{c}$ is the same number value in
           $\bar{C}^{c}_{x}$ as $a_{x}$ is in $\bar{C}_{x}.$ As shown by the
           lines they are values for different elements of $C_{x}.$ The
           lines also show that the $C_{x}$ element that has the value
           $a_{c}$ in $\bar{C}^{c}_{x}$, has the value $ca_{x}$
           in $\bar{C}_{x}.$ Subscripts denote structure
            memberships of the number values.}\label{NPIc1}\end{figure}

            Figure \ref{NPIc1} shows the relation between the valuation
            of elements of $C_{x}$ in $\bar{C}_{x}$ and in $\bar{C}^{c}_{x}.$
            The relations outlined above are shown by the arrows and
            number values in the figure.

            The one exception is the element of $C_{x}$ that has value $0$ in $\bar{C}_{x}.$ This value is the same in all $\bar{C}^{c}_{x}$ for all values of $c.$  In this sense it is "the number vacuum" in that  it is unchanged under all transformations,\footnote{Like the physical vacuum which is invariant under all space time translations.} $\bar{C}^{c}_{x}\rightarrow\bar{C}^{c'}_{x}.$

            The relations between the basic operations and numbers in $\bar{C}^{c}_{x}$ and those in $\bar{C}_{x}$ are not arbitrary.
            They are determined by the requirement that $\bar{C}_{x}$ satisfies the axioms\footnote{The axioms describe a smallest closed algebraic field of characteristic $0.$} for complex numbers \cite{complex,comcon} if and only if $\bar{C}^{c}_{x}$ does.
            It is shown elsewhere \cite{BenRENT,BenNGF} that this requirement is satisfied by the relations shown in Eq. \ref{Wcx}.

            The relations between number values in $\bar{C}^{c}_{x}$
            and those in $\bar{C}_{x}$ extend to terms and functions.
            Let $t_{c}$ be a term in $\bar{C}^{c}_{x}$ where \begin{equation}\label{tc} t_{c}=(\sum_{j,k=1}^{n,m})_{c} \frac{a_{c}^{j}}{b_{c}^{k}}.\end{equation}

            The corresponding term in $\bar{C}_{x}$ is
            obtained by replacing number values, and the implied sum,
            multiplications, and divisions in $\bar{C}^{c}_{x}$ by
            their representations in $\bar{C}_{x}$ as given in Eq.
            \ref{Wcx}. In this case the $j$ values and $j-1$
            multiplications in the numerator give a factor of
            $c^{j-(j-1)}=c.$  This is canceled by a similar $c$
            factor in the denominator. The solidus, as a division,
            contributes a factor of $c$ to give as a final result:
            \begin{equation}\label{tcx}t_{c}=(\sum_{j,k=1}^{n,m})_{c}
            \frac{a_{c}^{j}}{b_{c}^{k}}=c(\sum_{j,k=1}^{n,m})_{x}
            \frac{a_{x}^{j}}{b_{x}^{k}}=ct_{x}.\end{equation}Here
            $t_{x}$ is the same term in $\bar{C}_{x}$ as $t_{c}$ is
            in $\bar{C}^{c}_{x}.$

            This result extends term by term to convergent power
            series and thus to analytic functions \cite{Rudin}. If
            $f_{c}(a_{c})$ is an analytic function on
            $\bar{C}^{c}_{x},$ then the corresponding analytic
            function on $\bar{C}_{x}$ is given by $cf_{x}(a_{x}).$
            Here $f_{x}$ is the same function in
            $\bar{C}_{x}$ as $f_{c}$ is in $\bar{C}^{c}_{x}$
            in that $f_{x}(a_{x})$ is the same number value in
            $\bar{C}_{x}$ as $f_{c}(a_{c})$ is in $\bar{C}^{c}_{x}.$

            \section{Gauge Fields}\label{GF}
            As was noted earlier, the above description of the
            local representation of $\bar{C}_{y}$ on $\bar{C}_{x},$
            as in Eq. \ref{Wcx}, is valid for $y=x+\hat{\nu}dx$ a
            neighbor point of $x.$ One would like to extend the
            description of local representations of $\bar{C}_{y}$ on
            $\bar{C}_{x}$ for points $y$ distant from $x.$  Also
            one would like to be able to use the results obtained
            so far in gauge theories.

           These and other considerations suggest that one represent
           $c=c_{y,x}$ in terms of a complex valued gauge field
           $\vec{A}(x)+i\vec{B}(x):$\begin{equation}\label{GFcyx}
           c_{y,x}=e^{(\vec{A}(x)+i\vec{B}(x))\cdot\hat{\nu}dx}.
           \end{equation} Both $\vec{A}(x)$ and $\vec{B}(x)$ are
           real valued gauge fields with four space time components
           $A_{\mu}(x),B_{\mu}(x).$

           This can be used to give an alternate expression for the
           action of $W^{c}_{x}$ on $\bar{C}_{x}.$ For  number
           values one obtains from Eq. \ref{Wcx}\begin{equation}
           \label{Wcxax}W^{c}_{x}a_ {x}=e^{(\vec{A}(x)+i\vec{B}(x))
           \cdot\hat{\nu}dx}a_{x}.\end{equation}  This shows that
           $W^{c}_{x}$ can be considered to be an element of the
           gauge group, $GL(1,C).$
            To first order in small quantities, one has
            \begin{equation}\label{Wcx1stord}\begin{array}{l}
            W^{c}_{x}(a_{x})=(1+(\vec{A}(x)+i\vec{B}(x))\cdot
            \hat{\nu}dx)a_{x}\\\\\hspace{1.5cm}=(1+(A_{\mu}(x)+iB_{\mu}(x))
            dx^{\mu})a_{x}.\end{array}\end{equation} For $y$
            a neighbor point of $x$, the scale change
            factor relating a local representation of $\bar{C}_{y}$
            to $\bar{C}_{x}$ is given by\begin{equation}\label{SCFCTR}
            (\vec{A}(x)+i\vec{B}(x))\cdot\hat{\nu}dx.\end{equation}

            These results can be used to define  a covariant
            derivative of a complex number valued field, $\psi(x)$
            over space time. As is well known \cite{Montvay}, the
            usual derivative\begin{equation}\label{partialmux}
            \partial_{\mu,x}\psi=\frac{\psi(x+dx^{\mu})-\psi(x)}{dx^{\mu}}
            \end{equation} is not defined as $\psi(x+dx^{\mu})$ and $\psi(x)$
            are in different complex number structures.  Subtraction
            is defined only within structures, not between
            structures.

            This can be solved by replacing $\psi(x+dx^{\mu})$ with
            $\psi(x+dx^{\mu})_{x}=F^{-1}_{x+dx^{\mu},x}\psi(x+dx^{\mu})$ to obtain
            \begin{equation}\label{partialp}\partial^{\prime}_{\mu,x}
            \psi=\frac{\psi(x+dx^{\mu})_{x}-\psi(x)}{dx^{\mu}}.
            \end{equation}Here $F^{-1}_{x+dx^{\mu},x}$ is the parallel
            transform operator from $\bar{C}_{x+dx^{\mu}}$ to $\bar{C}_{x}$
            and $\psi(x+dx^{\mu})_{x}$ is the same number value in
            $\bar{C}_{x}$ as $\psi(x+dx^{\mu})$ is in $\bar{C}_{x+dx^{\mu}}.$

            However, this does not take into account the freedom of
            choice of scaling between $\bar{C}_{x+dx^{\mu}}$ and
            $\bar{C}_{x}.$ This extends to complex number structures,
            the freedom of basis choice in vector spaces that
            is used in gauge theories.

            Taking this into account gives the covariant derivative
            $D_{\mu,x}$ where\begin{equation}\label{Dmux}\begin{array}{l}
           D_{\mu,x}\psi=\frac{\textstyle e^{(A_{\mu}(x)+iB_{\mu}(x))dx^{\mu}}
           \psi(x+dx^{\mu})_{x}-\psi(x)}{\textstyle dx^{\mu}}
           \\\\\hspace{1.5cm}=
           \partial^{\prime}_{\mu,x}\psi+(A_{\mu}(x)+iB_{\mu}(x))
           \psi(x+dx^{\mu})_{x}.\end{array}\end{equation}The use
           of this in gauge theories will be discussed shortly.

           \subsection{Number Representation at Distant
           Points}\label{NRDP}
           So far the discussion has been pretty much limited to $y$
           a neighbor point of $x.$ It needs to be extended to cases
           where $y$ is distant from $x.$ First consider a two step
           path $x\rightarrow y\rightarrow z$ where $y=x+\hat{\nu}_{1}
           \Delta_{x}$ and $z=y+\hat{\nu}_{2}\Delta_{y}.$ $\Delta_{y}$
           and $\Delta_{x}$ are small distances with number values
           in $\bar{C}_{y}$ and $\bar{C}_{x}$ respectively.

           Let $a_{z}$ be a number value in $\bar{C}_{z}.$ The
           corresponding number value in $\bar{C}_{y}$ is
           $c_{z,y}\times_{y}a_{y}.$  Here $c_{z,y}$ is the complex
           scaling factor on the link from $y$ to $z$ and
           $a_{y}=F^{-1}_{z,y}a_{z}$ is the same number  value in
           $\bar{C}_{y}$ as $a_{z}$ is in $\bar{C}_{z}.$

           The number value in $\bar{C}_{x}$ that corresponds to
           $c_{z,y}\times_{y}a_{y}$ in $\bar{C}_{y}$ is given by
           \begin{equation}\label{2stpP}c_{y,x}(c_{z,y})_{x}
           \frac{\times_{x}}{c_{y,x}}c_{y,x}a_{x}=c_{y,x}
           (c_{z,y})_{x}a_{x}.\end{equation} Here $(c_{z,y})_{x}
           =F^{-1}_{y,x}c_{z,y}$ and $a_{x}=F^{-1}_{y,x}a_{y}$
           are the same number values in $\bar{C}_{x}$ as
           $c_{z,y}$ and $a_{y}$ are in $\bar{C}_{y}.$

           An expression equivalent to Eq.\ref{2stpP} can be
           obtained by use of Eq. \ref{Wcx1stord}. To first
           order one obtains\begin{equation}\label{czyx1st}
           \begin{array}{l}c_{y,x}(c_{z,y})_{x}a_{x}=[1+
           (\vec{A}(x)+i\vec{B}(x))\cdot\hat{\nu}_{1}\Delta_{x}
           +(\vec{A}(y)_{x}+i\vec{B}(y)_{x})\cdot\hat{\nu}_{2}
           (\Delta_{y})_{x}]a_{x}\\\\\hspace{2cm}=[1+(\vec{A}(x)+i\vec{B}(x))
           \cdot\hat{\nu}_{1}+(\vec{A}(y)_{x}+i\vec{B}(y)_{x})
           \cdot\hat{\nu}_{2}]\Delta_{x}a_{x}.\end{array}
           \end{equation}

           Here $\vec{A}(y)_{x}$ and $\vec{B}(y)_{x}$ are the same real
           valued vectors at $x$ as they are at $y,$\footnote{$\vec{A}(y)_{x}$
           can be expressed as the parallel transform,
           $F^{-1}_{y,x}A_{\mu}(y)=A_{\mu}(y)_{x},$  of the components,
           $A_{\mu}(y)$ which are real values in $\bar{C}_{y}$,
           to $\bar{C}_{x}.$  The same argument holds for $\vec{B}.$} and
           $(\Delta_{y})_{x}$ has been set equal to $\Delta_{x}.$

           The extension to an $n$ step path is straight forward.
           Let $P$ be an $n$ step path where $P(0)=x_{0}=x,
           P(j)=x_{j}, P(n-1)=x_{n-1}=y$ and
           $x_{j+1}=x_{j}+\hat{\nu}_{j}\Delta_{x_{j}}.$  Then
           \begin{equation}\label{WyPx}W^{y,P}_{x}a_{x}=
           c^{P}_{y,x}a_{x}\end{equation} is the local
           representation of $a_{y}$ in $\bar{C}_{x}.$  Here
           \begin{equation}\label{cPyx}
           c^{P}_{y,x}=\prod_{j=0}^{n-1}(c_{x_{j+1},x_{j}})_{x}
           =\exp(\sum_{j=0}^{n-1}[(\vec{A}(x_{j})+i\vec{B}(x_{j}))
           \cdot\hat{\nu}_{j}\Delta_{x_{j}}]_{x}).\end{equation} The
           subscript $x$ denotes the fact that all terms in the sum,
           the sum, and the exponential, are values in
           $\bar{C}_{x}.$ An ordering of terms in the product of Eq,
           \ref{cPyx} is not needed because the different $c$
           factors commute with one another.

           Let $P$ be a continuous path with points parameterized by
           $s$.  $s$ is a continuous variable from $0$ to $1$ with
           $P(0)=x,P(1)=y.$ $c^{P}_{y,x}$ can be expressed in terms
           of the gauge fields \cite{BenNGF} by \begin{equation}
           \label{cPyxint} c^{P}_{y,x}=\exp\{\int_{0}^{1}
           (\vec{A}(P(s))_{x}+i\vec{B}(P(s))_{x})\cdot
           [\frac{dP(s)} {d s}]_{x}ds\}.\end{equation} The
           derivative components, $[\frac{dP(s)} {d s}]_{x},$ are
           the same number values in $\bar{C}_{x}$ as  the
           $\frac{dP(s)}{ds}$ are in $\bar{C}_{P(s)}.$

           An equivalent expression for $c^{P}_{y,x}$ is as a line
           integral along the path: \begin{equation}\label{LIcPyx}
           c^{P}_{y,x}=\exp(\int_{P}(\vec{A}(z)_{x}+i\vec{B}(z)_{x})\vec{dz}).
           \end{equation}The subscript $x$ indicates that the integral
           is defined in $\bar{C}_{x}.$

           \subsection{Space Integrals}\label{SI}
           The presence of the gauge fields affects space integrals
           of fields. As a simple example, let $\Phi(x)$ be a
           field where for each $x,$ $\Phi(x)$ is a number value in
           $\bar{C}_{x}.$  The integral, $\int\Phi(y)dy$ is supposed
           to be the limit of a sum $\sum_{y}\Phi(y)\Delta_{y}$ as the
           cubic volume elements $\Delta_{y}\rightarrow 0.$

           The problem is that the sum is not defined as the
           elements of the sum are in different complex number
           structures and addition is not defined between
           structures. One way to fix this is to select a reference
           complex number structure, $\bar{C}_{x},$ and parallel
           transform the elements of the sum to $\bar{C}_{x}$ and
           then perform the summation and limit.  This would give,
           \begin{equation}\label{IPhiy}\int_{x}\Phi(y)dy=
           \lim_{\Delta_{x}\rightarrow 0}\sum_{y}F^{-1}_{y,x}
           (\Phi(y)\Delta_{y})=\lim_{\Delta_{x}\rightarrow
           0}\sum_{y}\Phi(y)_{x}\Delta_{x}.\end{equation} The
           subscript, $x$ on $\int$ indicates that the integral is
           defined on $\bar{C}_{x}.$ Also $\Phi(y)_{x}$ and
           $\Delta_{x}$ are the same number values in $\bar{C}_{x}$
           as $\Phi(y)$ and $\Delta_{y}$ are in $\bar{C}_{y}.$

           However, this representation of $\int\Phi(y)dy$ does not
           include the freedom of choice of scale factors.
           Inclusion of this freedom into the expression for the
           integral gives\begin{equation}\label{IPhiyP}\int_{x,P}\Phi(y)dy=
           =\lim_{\Delta_{x}\rightarrow 0}\sum_{y}c^{P}_{y,x}\Phi(y)_{x}
           \Delta_{x}=\int c^{P}_{y,x}\Phi(y)_{x}dy_{x}.\end{equation}
           Here $c^{P}_{y,x}$ is given by Eq. \ref{cPyx}.

           The problem here is the dependence of the integral on the
           path $P$ from $x$ to $y.$ This would introduce serious
           problems into the definitions of these integrals as
           one would have to define some sort of path integral.

           This problem can be avoided if
           the gauge fields $\vec{A}(x)$ and $\vec{B}(x)$ are
           integrable.\footnote{Integrals of the fields from $x$
           to $y$ are independent of the path chosen.} In this case $c^{P}_{y,x}$ is independent of
           $P$ and depends on $x$ and $y$ only. Then\begin{equation}
           \label{IcPyx}\int c^{P}_{y,x}\Phi(y)_{x}dy_{x}=\int
           c_{y,x}\Phi(y)_{x}dy_{x}\end{equation} where\begin{equation}
           \label{intcyx}c_{y,x}=\exp(\int(\vec{A}(z)_{x}+i\vec{B}
           (z)_{x})dz)\end{equation}

            At present it is not known if either $\vec{A}$ or
           $\vec{B}$ are integrable or not.  Future work should help
           to decide this question.

           \section{Other Mathematical Systems}\label{OMS}
           So far the effect of choice freedom of scaling factors
           has been limited to complex number structures.  One would
           expect it to also effect other mathematical systems that
           are based on numbers.  Vector spaces are examples as they
           are closed under scalar vector multiplication. If they
           are normed spaces, then the norms are scalars.

           Here Hilbert spaces are considered as examples of
           vector spaces. As noted in the introduction, the setup
           considered here consists of an assignment of a Hilbert
           space structure and a complex number structure,
           $\bar{H}_{x},\bar{C}_{x}$ to each space time point.
           $\bar{C}_{x}$ is the set of scalars for $\bar{H}_{x}.$

           $\bar{H}_{x}$ and $\bar{H}_{y}$ are given by
           \begin{equation}\label{barHxy}\begin{array}{c}
           \bar{H}_{x}=\{H_{x},+_{x},-_{x},\cdot_{x}, \langle
           -,-\rangle_{x},\psi_{x}\}\\\bar{H}_{x}=\{H_{x},+_{x},
           -_{x},\cdot_{x}, \langle-,-\rangle_{x},\psi_{x}\}.\end{array}
           \end{equation}  $H_{x}$  and $H_{y}$ denote  base sets,
           $\cdot$ and $+,-$ denote scalar vector multiplication and
           linear superposition, and $\langle -,-\rangle$ denotes
           scalar product. The subscripts $x,y$ denote structure
           membership. Also $\psi_{y}$ is the same vector value in
           $\bar{H}_{y}$ and $\psi_{x}$ is in $\bar{H}_{x}.$
           $\psi_{x},\psi_{x}$ are to be distinguished from
           $\psi(y)$ which is a field.\footnote{The basic operations
           shown in Eq. \ref{barHxy} must satisfy
           the axioms for a Hilbert space. These describe a complex
           inner product vector space that is complete in the
           norm \cite{Kadison1}.}

           As was noted in Section \ref{FPT} $\bar{H}_{y}$ and
           $\bar{H}_{x}$ are related by a  unitary parallel transform
           operator $U_{y,x}$ where $\bar{H}_{y}=U_{y,x}\bar{H}_{x}.$
           If $\psi_{x}$ is a vector value in $\bar{H}_{x}$, then
           $\psi_{y}=U_{y,x}\psi_{x}$ is the same vector value in
           $\bar{H}_{y}$ as $\psi_{x}$ is in $\bar{H}_{x}.$

           The freedom of basis choice \cite{Yang} in gauge theories
           \cite{Montvay}, applied here requires the factorization
           of $U_{y,x}$ into two factors as in Eq. \ref{UXV}.  This
           can be used to define a local representation, $\bar{H}^{V}_{x},$
           of $\bar{H}_{y}$ on $\bar{H}_{x}$ by\begin{equation}\label{HVx}
           \bar{H}^{V}_{x}=\{H_{x},\pm_{x}\cdot_{x},\langle-,-\rangle_{x},
           V_{y,x}\psi_{x}\}.\end{equation} Here\begin{equation}
           \label{Vyxpsix}V_{y,x}\psi_{x}=V_{y,x}U^{\dag}_{y,x}
           \psi_{y}=X^{\dag}_{y,x}\psi_{y}\end{equation}is the local representation of $\psi_{y}$ at $x.$

           This takes account of the freedom of basis choice but not
           the freedom of scaling choice for the scalar fields. This
           can be accounted for by defining the Hilbert space structure,
           $\bar{H}^{cV}_{x},$ for which $\bar{C}^{c}_{x},$ Eq. \ref{bCcx},
           is the scalar field structure.  Here \begin{equation}\label{HcVx}
           \bar{H}^{cV}_{x}=\{H_{x},\pm^{cV}_{x},\cdot^{cV}_{x},\langle
           -,-\rangle^{cV}_{x},\psi^{cV}_{x}\}\end{equation} is the
           local representation of $\bar{H}_{y}$ at $x.$

           As was the case for complex number structures one needs
           to give a specific representation of the operations and
           vector values of $\bar{H}^{cV}_{x}$ in terms of those of
           $\bar{H}_{x}.$  These are given by another representation
           of $\bar{H}^{cV}_{x}$ as\begin{equation}\label{HcVx1}
           \bar{H}^{cV}_{x}=\{H_{x},\pm_{x},\frac{\cdot_{x}}{c},
           \frac{\langle -,- \rangle_{x}}{c^{*_{x}}},cV\psi_{x}\}.
           \end{equation}This representation of $\bar{H}^{cV}_{x}$
           is referred to as the local representation of
           $\bar{H}_{y}$ on $\bar{H}_{x}.$ The scalar field for this representation is $\bar{C}^{c}_{x},$ given by Eq. \ref{Ccx}.

           It follows that the local representation, in $\bar{H}_{x},$
           of a vector $\psi(y)$ in $\bar{H}_{y}$ is given by\begin{equation}\label{Xdpsiy}X^{\dag}_{y,x}\psi(y) =c_{y,x}V_{y,x}\psi(y)_{x}.\end{equation} Here $\psi(y)_{x}
           =U^{-1}_{y,x}\psi(y)$ is the same vector in $\bar{H}_{x}$
           as $\psi(y)$ is in $\bar{H}_{y}.$

           The appearance of $c$ in the denominator of the scalar
           vector multiplication follows from the following equivalences:
           \begin{equation}\label{eqcdot}\begin{array}{l}\phi^{cV}_{x}=a_{c}
           \cdot^{cV}_{x}\psi^{cV}_{x}\Leftrightarrow cV\phi_{x}
           =(ca_{x})\;\frac{\textstyle \cdot_{x}}{\textstyle c}\; cV\psi_{x}
           \\\\\hspace{1cm}\Leftrightarrow cV\phi_{x}=(ca_{x})\cdot_{x} V
           \psi_{x}\Leftrightarrow\phi_{x}=a_{x}\psi_{x}.\end{array}
           \end{equation}These show that, as required, $\phi^{cV}_{x}=
           a_{c}\cdot^{cV}_{x}\psi^{cV}_{x}$ is true in $\bar{H}^{cV}_{x}$ if
           and only if $\phi_{x}=a_{x}\cdot_{x}\psi_{x}$ is true in $\bar{H}_{x}.$

           For the scalar product in Eq. \ref{HcVx1} the equivalences are:
           \begin{equation}\label{eqscpr}\begin{array}{l}
           a_{c}=\langle\phi^{cV}_{x},\psi^{cV}_{x}\rangle^{cV}_{x}
           \Leftrightarrow ca_{x}=\langle cV\phi_{x},
           cV\psi_{x}\rangle_{c}\\\\\hspace{1cm}\Leftrightarrow ca_{x}=c\langle V\phi_{x},
           V\psi_{x}\rangle_{x}\Leftrightarrow a_{x}=\langle \phi_{x},
           \psi_{x}\rangle_{x}.\end{array}\end{equation} If $|cV\phi_{x}\rangle_{c}$ in $\bar{H}^{cV}_{x}$ becomes $cV|\phi_{x}\rangle_{x}$ in $\bar{H}_{x},$ then
           \begin{equation}\label{eqsp}ca_{x}= \langle cV\phi_{x},cV\psi_{x}\rangle_{c}\Leftrightarrow ca_{x}=c^{*_{x}}c\frac{\textstyle\langle V\phi_{x},
           V\psi_{x}\rangle_{x}}{\textstyle c^{*_{x}}}\Leftrightarrow a_{x}=\langle \phi_{x},
           \psi_{x}\rangle_{x}.\end{equation} 
           
           Eqs. \ref{eqscpr} and \ref{eqsp}
           show that $a_{c}=\langle\phi^{cV}_{x},\psi^{cV}_{x}\rangle^{cV}_{x}$ is true
           in $\bar{H}^{cV}_{x}$ and $\bar{C}^{c}_{x}$ if and only if $a_{x}=
           \langle\phi_{x},\psi_{x}\rangle_{x}$ is true in $\bar{H}_{x}$
           and $\bar{C}_{x}.$  These equations also show the reason for $c^{*_{x}}$ as a scalar product divisor in Eq. \ref{HcVx1}

           One may wonder if the presence of $c,$ as a factor multiplying
           $V\psi$  in Eq. \ref{HcVx1}, is needed. It is needed if one accepts
           the equivalence $\bar{H}\simeq\bar{C}^{n}$ \cite{Kadison} between
           $n$ dimensional Hilbert spaces and complex number tuples. Use of
           this for each   point, $x,$ gives $\bar{H}_{y}\simeq\bar{C}^{n}_{y}.$
           Similarly $\bar{H}^{cV}_{x}\simeq(\bar{C}^{c}_{x})^{n}.$
           Eqs. \ref{HcVx1} and \ref{Ccx} are used here.

           To examine in more detail, it is sufficient to set $V=1.$ A
           vector in $(\bar{C}^{c}_{x})^{n}$ is a column of $n$
           complex numbers, $a_{c,i}:i=1,\cdots,n.$  The
           corresponding column in $\bar{C}_{x}$ is
           $ca_{x,i}:i=1,\cdots,n.$ It follows that any vector
           $\psi^{c}_{x}$ in $\bar{H}^{c}_{x}$ corresponds
           to $c\psi_{x}$ in $\bar{H}_{x}.$

           The presence of the gauge fields affects derivatives of fields.
           Let $\psi$ be a matter field where $\psi(x)$ is an element of
           $\bar{H}_{x}.$ The usual derivative \begin{equation}\label{parmux}
           \partial_{\mu,x}\psi=\frac{\psi(x+dx^{\mu})-\psi(x)}{\partial x^{\mu}}
           \end{equation} does not make sense because subtraction is not
           defined between different vector spaces.

           One way to cure this is to replace  $\partial_{\mu,x}$ with
           $\partial^{\prime}_{\mu,x}$ where\begin{equation} \label{parpmux}
           \partial^{\prime}_{\mu,x}\psi=\frac{\psi(x+dx^{\mu})_{x}
           -\psi(x)}{\partial x^{\mu}}\end{equation} Here
           $\psi(x+dx^{\mu})_{x} =U^{-1}_{x+dx^{\mu},x}
           \psi(x+dx^{\mu})$ is the same vector in $\bar{H}_{x}$ as $\psi(x+dx^{\mu})$ is in $\bar{H}_{x+dx^{\mu}}.$

           However, this does not take account of the freedom of choice of
           scaling introduced here or the freedom of basis choice. This is
           accounted for  by replacing $\psi(x+dx^{\mu})_{x}$ by the local
           representation of $\psi(x+dx^{\mu})$ in $\bar{H}_{x}$ as given
           by Eq. \ref{Xdpsiy}. This gives the expression for the covariant derivative
           \begin{equation}\label{DmuxcV}D_{\mu,x}\psi=\frac{c_{x+dx^{\mu},
           x}V_{x+dx^{\mu},x}\psi(x+dx^{\mu})_{x}-\psi(x)}{\partial x^{\mu}}.
           \end{equation}

           \section{Gauge Theories}\label{GT}
           As is well known, physical Lagrangians include a covariant
           derivative. Examples include the Klein Gordon  and Dirac
           Lagrangians: \begin{equation}\label{KleinG} \mathcal{L}(x)=\psi^{\dag}(x)D^{\mu}_{x}D_{\mu,x}\psi -m^{2} \bar{\psi}(x)\psi(x) \end{equation} and\begin{equation}\label{Dirac}\mathcal{L}(x)=
           \bar{\psi}(x)i\gamma^{\mu}D_{\mu,x}\psi-m\bar{\psi}(x)\psi(x).
           \end{equation} The usual treatment uses Eq. \ref{DmuxcV} as an
           expression for the covariant derivative with $c_{y,x}=1$
           everywhere.  As such it is a special case of the setup described here.

           Inclusion of the freedom of scaling choice described here
           results in use of Eq. \ref{DmuxcV} for the covariant
           derivative in Lagrangians. In this case the usual gauge
           group for an $n$ dimensional vector space space is expanded
           from $U(n)$ to $GL(1,C)\times SU(n).$ Here $c_{x+dx^{\mu},x}$
           belongs to $GL(1,C)$ and $V_{x+dx^{\mu},x}$ belongs to  $SU(n).$
           Note that the $U(1)$ factor of $U(n)$ is not present as it is already included in $GL(1,C).$ This will be discussed more later on.

           The replacement of $U(n)$ by $GL(1,C)\times SU(n)$ has
           consequences for both Abelian and nonabelian gauge theories.
           For Abelian theories the gauge group is $GL(1,C)$.  For these
           theories, replacement of $c_{x+dx^{\mu},x}$ by its Lie algebra
           representation, Eq.\ref{GFcyx}, and expansion to first order
           gives Eq. \ref{Dmux} which is repeated here:\begin{equation}
           \label{1Dmux}D_{\mu,x}\psi=\partial^{\prime}_{\mu,x}\psi+
           (g_{R}A_{\mu}(x)+ig_{I}B_{\mu}(x))\psi(x+dx^{\mu})_{x}.
           \end{equation} Coupling constants $g_{R}$ and $g_{I}$ have
           been added to the $\vec{A}(x)$ and $\vec{B}(x)$ fields.

           One now imposes the requirement that terms in  the Lagrangians
           are limited to those that are invariant under global and local
           gauge transformations \cite{Novaes}. For Abelian theories, global gauge
           transformations have the form \begin{equation}\label{GlGT}
           \Lambda =e^{i\phi}\end{equation} where $\Lambda$ is a constant.
           Nonlocal gauge transformations have the form \begin{equation}\label{LcGT}
           \Lambda(x)=e^{i\phi(x)}\end{equation} where $\Lambda(x)$ depends on $x$ through the $x$ dependence of $\phi(x).$

           One replaces $\psi(x)$ in the Lagrangians with $\psi'(x)=
           \Lambda(x)\psi(x)$ and examines the terms for invariance. Since
           $(\psi')^{\dag}(x)\psi'(x)=\psi^{\dag}\psi(x),$ terms of this form remain. For terms involving the derivative one follows the standard procedure \cite{Cheng}. Invariance under local gauge transformations requires that \begin{equation}\label{DpmuxL}
           D^{\prime}_{\mu,x}\Lambda\psi=\Lambda(x)D_{\mu,x}\psi
           \end{equation} hold. $D^{\prime}_{\mu,x}$ is given by Eq. \ref{1Dmux} where $A^{\prime}_{\mu}(x)$ and $B^{\prime}_{\mu}(x)$ replace $A_{\mu}(x)$ and $B_{\mu}(x).$

           Solving this equation for $A^{\prime}_{\mu}(x)$ and $B^{\prime}_{\mu}(x)$ as a function of $A_{\mu}(x),B_{\mu}(x)$ and $\Lambda(x)$ gives \begin{equation}\label{ApBp}
           \begin{array}{c}A^{\prime}_{\mu}(x)=A_{\mu}(x)
           \\\\B^{\prime}_{\mu}(x)=B_{\mu}(x)+\frac{\textstyle i\Lambda^{-1}(x)\partial^{\prime}_{\mu,x}\Lambda}{\textstyle g_{I}}=B_{\mu}(x)-\frac{\textstyle \partial^{\prime}_{\mu,x}\phi(x)}{\textstyle g_{I}}\end{array}
           \end{equation}

           These results show that $\vec{A}(x)$ is gauge invariant and that $\vec{B}(x)$ depends on the local gauge transformation. It follows from this that $\vec{A}(x)$ and $\vec{B}(x)$ correspond to two gauge bosons, $\vec{A}(x)$ can have mass in the sense that a mass term is optional in the lagrangian. However $\vec{B}(x)$ must be massless.  The reason is that a mass term $\vec{B}^{\dag}(x)\vec{B}(x)$ is not local gauge invariant.

           The dynamics of the massless boson can be added to a Lagrangian by a Yang Mills term \begin{equation}\label{GG} \frac{1}{4}G_{\mu,\nu}(x)G^{\mu,\nu}(x)\end{equation} where
           \begin{equation}\label{Gmn}G_{\mu,\nu}=
           \partial^{\prime}_{\mu,x}\vec{B}_{\nu}(x)-
           \partial^{\prime}_{\nu,x}B_{\mu}(x).\end{equation}Addition of
           $1/4G_{\mu,\nu}G^{\mu,\nu}$ and a mass term for the
           $\vec{A}(x)$ field to the Dirac Lagrangian gives, \begin{equation}\label{Diracexp}\begin{array}{l}L(x)=
          \bar{\psi}i\gamma^{\mu}(\partial^{\prime}_{\mu,x}+g_{R}A_{\mu}(x)
          +ig_{I}B_{\mu}(x))\psi-m\bar{\psi}\psi\\\\\hspace{1cm}-
          \frac{1}{2}\lambda^{2}A^{\mu}(x)A_{\mu}(x)-\frac{1}{4}
          G_{\mu,\nu}G^{\mu,\nu}.\end{array}\end{equation} This is equivalent to the QED Lagrangian with  additional terms for the $\vec{A}(x)$ field.

          For nonabelian gauge theories, such as that for the gauge group $GL(1,C)\times SU(2)$ there is another equation added to Eq. \ref{ApBp} for the vector bosons.  Since there is no change in the first two equations, the results for the vector bosons are not relevant here.  A brief summary is in \cite{BenNGF}.

          \section{Discussion}\label{Di}
          There are some questions and problems that arise with the number scaling introduced here. The main one regards the physical nature, if any, of the $\vec{A}(x)$ and $\vec{B}(x)$ fields.

          The fact that setting  $\vec{A}(x)=0$ in the Dirac Lagrangian gives the QED Lagrangian suggests strongly that $\vec{B}(x)$ is the photon or electromagnetic field. In this case the coupling constant $g_{I}=e$ where $e$ is the electric charge. It is important to note that this assignment is based on the complete suppression of the $U(1)$ component of the vector space gauge group. The reason is that as far as the mathematics is concerned it contributes a gauge field that is identical to $\vec{B}(x).$ This can be seen by expanding the overall gauge group to be $GL(1,C)\times U(n)$ and carrying out the usual gauge theory treatment. If one assigns the same coupling constant to both field components, $\vec{\Gamma}(x),$ from $U(1)$ and $\vec{B}(x),$ then only the sum, $\vec{B}(x)+\vec{\Gamma}(x),$ of the fields appears in the Lagrangian. Here the $U(1)$ component is given by $\exp(i\vec{\Gamma}(x)).$

          It is an open question whether the photon field is just $\vec{B}(x),\vec{\Gamma}(x),$ or a combination of both. The presence of both fields is  is not likely as can be seen by considering the equivalence $\bar{H}_{x}\simeq\bar{C}^{n}_{x}.$ In this case the representation, $\bar{H}^{cV}_{x},$ Eq. \ref{HcVx1}, of $\bar{H}_{y}$ on $\bar{H}_{x}$ along with the representation, of $\bar{C}_{y}$  on $\bar{C}_{x},$ gives $\bar{H}^{cV}_{x}\simeq V_{y,x}(C^{c}_{x})^{n}.$ Here $\bar{C}^{c}_{x}$ is given by Eq. \ref{Ccx}.

          This representation argues for assigning $\vec{B}(x)$ to be the photon and  $V_{y,x}$ to belong to $SU(n).$ One reason is that the Hilbert space representation is constructed from $\bar{C}^{c}_{x}$ which already has the $\vec{B}(x)$ field present. However, more work is needed here.

          The physical nature of the real $\vec{A}(x)$ field is open. Candidate fields include the Higgs boson, dark matter, dark energy, and gravity. One feature that may help decide is that the coupling constant, $g_{R},$ of $\vec{A}(x)$ to  matter fields must be very small compared to the fine structure constant. This is based on the great accuracy of QED.

          Finally it is worth noting that the space time scaling of complex, (and other types \cite{BenRENT} of) numbers described here may be a good approach to developing a coherent theory of physics and mathematics together \cite{BenTCTPM,BenTACTPM}. Clearly there is much work to do.

            \section*{Acknowledgement}
          This work was supported by the U.S. Department of Energy,
          Office of Nuclear Physics, under Contract No.
          DE-AC02-06CH11357.


\begin{thebibliography}{99}

            \bibitem{Yang}
            C. N. Yang and R. L. Mills, Phys. Rev. \textbf{96}, 191
            (1954).

                \bibitem{Novaes}
           S. F. Novaes, "Particles and Fields", Proceedings, X
           Jorge Andre Swieca Summer School, Sao Paulo,
           February 1999, Editors, J. Barata, M. Begalli,
           R. Rosenfeld, World Scientific Publishing Co.
           Singapore, 2000; Arxiv:hep-th/0001283.

            \bibitem{Mack}
            G. Mack, Fortshritte der Physik, \textbf{29}, 135
            (1981).

            \bibitem{Montvay}
            I. Montvay and G. M\"{u}nster, \emph{Quantum Fields on a
            Lattice}, Cambridge Monographs on Mathematical Physics,
            Cambridge University Press, UK, 1994.

            \bibitem{BenRENT}
            P. Benioff, arXiv 1102.3658.

            \bibitem{BenNGF}
            P. Benioff, arXiv 1008.3134.

            \bibitem{Barwise}
            J. Barwise, An Introduction to First Order Logic, in
            \emph{Handbook of Mathematical Logic}, J. Barwise, Ed.
            North-Holland Publishing Co. New York, 1977. pp 5-46.

            \bibitem{Keisler}
            H. J. Keisler, Fundamentals of Model Theory, in
            \emph{Handbook of Mathematical Logic}, J. Barwise, Ed.
            North-Holland Publishing Co. New York, 1977. pp 47-104.

            \bibitem{complex}
            J. Shoenfield, \emph{Mathematical Logic}, Addison Weseley
            Publishing Co. Inc. Reading Ma, 1967, p. 86; Wikipedia:
            Complex Numbers.

            \bibitem{Kaye}
            R.Kaye, \emph{Models of Peano Arithmetic} Clarendon Press, Oxford,
            1991, pp 16-21.

            \bibitem{comcon}
            Wikipedia: Complex Conjugate.

              \bibitem{Rudin}
            W. Rudin, \emph{Principles of Mathematical Analysis},
            3rd Edition, McGraw Hill Inc. New York, 1976 Chapter 1,
            "The real and complex numbers". (Wikipedia: Real
            Numbers)

            \bibitem{Kadison1}
           R. V. Kadison and J. R, Ringrose, \emph{Fundamentals of
           the Theory of Operator Algebras: Elementary theory},
           Academic Press, New York, (1983), Chap 2.


            \bibitem{Kadison}
           R. V. Kadison and J. R, Ringrose, \emph{Fundamentals of
           the Theory of Operator Algebras: Elementary theory},
           Academic Press, New York, (1983), Chap 2, p. 83.

           \bibitem{Cheng}
           T. P. Cheng and L. F. Li, \emph{Gauge Theory of Elementary  Particle Physics}, Oxford University Press, Oxford, UK, (1984), Chapter 8.

            \bibitem{BenTCTPM}
            P. Benioff, Found. Phys. \textbf{35}, 1825-1856, (2005),
            Arxiv:quant-ph/0403209.

            \bibitem{BenTACTPM}
            P. Benioff, Found. Phys. \textbf{32},  989-1029, (2002),
            Arxiv:quant-ph/0201093.

            \end{thebibliography}
            \end{document}